\documentclass[a4paper]{spie}  
\usepackage[]{graphicx}
\usepackage{amssymb}

\def\eq#1{{(\ref{#1})}}

\title{Estimation of the concentration of deep traps in organic photoconductors using two-photon absorption}

\author{S.V. Novikov\supit{a}, A.R. Tameev\supit{a}, A.V. Vannikov\supit{a},  and J.-M. Nunzi\supit{b}
\skiplinehalf
\supit{a}A.N. Frumkin Institute of Physical Chemistry and Electrochemistry, Leninsky prosp., 31, Moscow 119991, Russia; \\
\supit{b}Queen's University, Kingston ON, K7L 3N6 Canada
}

\authorinfo{Further author information: (Send correspondence to S.V.N.)\\S.V.N.: E-mail: cnovikov@gmail.com, Telephone: +7 495 952 2428\\A.R.T.: E-mail: a.tameev@gmail.com, Telephone: +7 495 952 2428}

\begin{document}
  \maketitle

\begin{abstract}
Typically, amorphous organic materials contain high density of traps. Traps hinder charge transport and, hence, affect various working parameters of organic electronic devices. In this paper we suggest a simple but reliable method for the estimation of the concentration of deep traps (traps that keep carriers for a time much longer than the typical transport time of the device). The method is based on the measurement of the dependence of the total charge, collected at the electrode, on the total initial charge, uniformly generated in the transport layer under the action of a light pulse. Advantages and limitations of the method are discussed and an experimental example of the estimation of the density of deep traps in photoconductive organic material  poly(2-methoxy-5-(2'-ethylhexyloxy)-1,4-phenylenevinylene (MEH-PPV) is provided.
\end{abstract}


\keywords{Organic materials, hopping charge transport, traps}

\section{introduction}
\label{sec:intro}

Most organic materials used in nowaday organic electronic devices are amorphous materials (organic glasses).\cite{Borsenberger:book} For that reason traps are ubiquitous in solid organic semiconductors. They could be chemical impurities, which are intrinsic to the organic material (in many devices organic polymers are used and polymers are notoriously difficult to purify) or accumulated as  products of the degradation of the organic material during the device operation, or they could be structural imperfections of the material; the possible nature of traps is very divers. From the point of view of the operation of electronic organic devices the most important and common characteristic of traps is their ability to capture charge carriers and keep them for a long time, thus hindering the charge transport and leading to the degradation of the performance of the devices (solar cells, light emitting diodes etc.) \cite{Borsenberger:book,Pope:book}. In some situations traps significantly modify transport characteristics of organic materials, such as the dependence of the carrier drift mobility $\mu$ on the applied electric field. \cite{veres,spie,jist} The most dangerous type of traps is the deep traps capable to keep the captured carriers on the time scale much longer than the relevant time scales of the device (i.e., infinitely). In some cases traps are favorable: consider, for example, a photorefractive process, where one of the major steps is the trapping of carriers of one particular sign \cite{PR}. In this report we suggest a method to estimate the concentration of deep traps in photoconductive organic materials and provide an experimental example of the estimation of the density of deep traps in the  typical organic transport material MEH-PPV.

\section{EXPERIMENTAL AND RESULTS}
\label{sec:exp}

In the experiments, two-photon absorption of the laser beam with the wavelength of 1064 nm in a MEH-PPV film has been used for the generation of electron-hole pairs. Weak absorption of the light in the film guaranties a uniform spatial distribution of generated charges. Electron-hole pairs were dissociated by the applied electric field $F$ and then drifted charges were collected at the electrodes.

\begin{figure}[thbp]
\begin{minipage}[c]{0.57\linewidth}
\begin{center}
\includegraphics[width=3.7in]{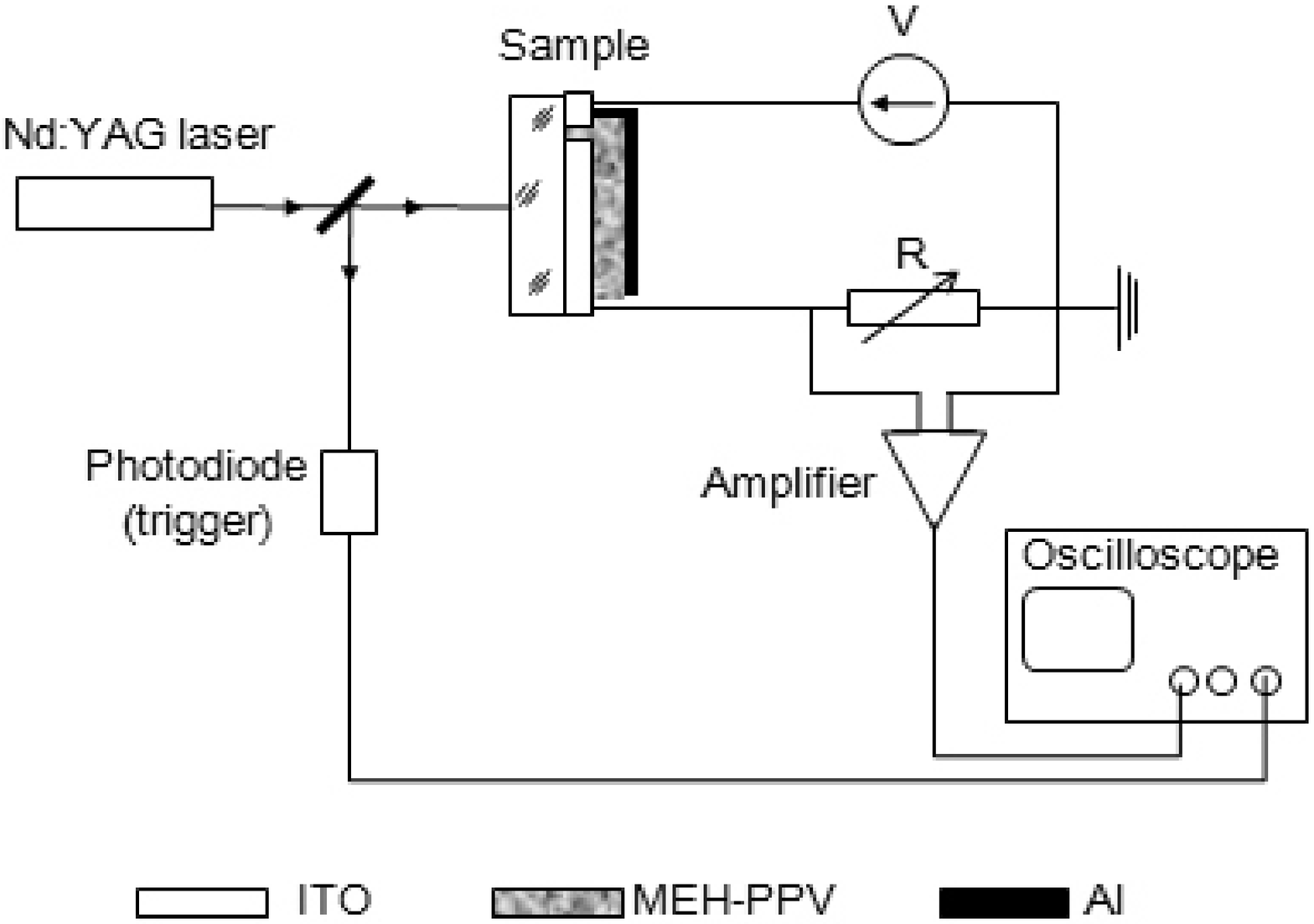}
\medskip
\newline (a)
\medskip
\end{center}
\end{minipage}
\begin{minipage}[c]{0.43\linewidth}
\begin{center}
\includegraphics[width=2.8in]{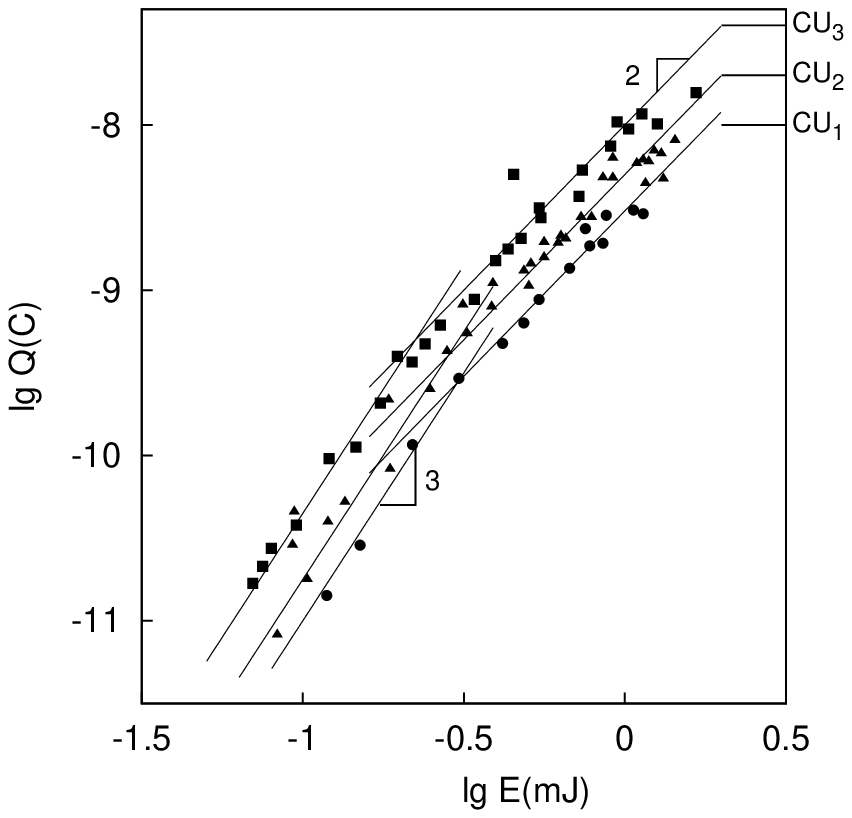}
\newline (b)
\medskip
\end{center}
\end{minipage}
\caption{(a) Experimental set-up. (b) Extracted charge $Q$ versus light pulse energy $E$ for three applied voltages $U$: 20V($\bullet$), 40V($\blacktriangle$), and 80V($\blacksquare$), correspondingly. Corresponding values of the electrode charge $CU$ are indicated at the right side of the plot. Capacitance of the specimen was $C=500$ pF. For this particular device the relation $m_0({\rm cm}^{-3})=1.25\times 10^{23}Q_t$(C) is valid.\label{Q-E}}
\end{figure}

The test specimens consisted of a layer of MEH-PPV sandwiched between the ITO and top aluminum electrodes. MEH-PPV from Aldrich was dissolved in toluene. The polymer film was deposited onto ITO/glass substrate by the doctor blade technique in air, and then dried in Ar atmosphere during 10 hours at 80$^{\rm o}$C. The thickness of the films varied between 1.5 $\mu$m and 2.0 $\mu$m. The standard Nd:YAG laser (pulse halfwidth of 25 ps) was used as a light source and the electric signals were recorded by Tektronix TDS3032 oscilloscope. The experimental set-up is sketched in Fig. \ref{Q-E}a.

Typical dependences of the total extracted charge $Q$ on the light pulse energy $E$ are shown in Fig. \ref{Q-E}b. For large $E$, where all deep traps are filled, $Q$ is approximately equal to the total generated charge $Q_0\propto E^2$, hence, $Q\propto E^2$. Indeed, the slope of $\log Q$ - $\log E$ plot is very close to 2.\footnote{For a comparison, in the case of one-photon excitation, where the light pulse is mostly absorbed in a thin surface layer, the dependence of $Q(E)$ is much weaker than $Q\propto E$ for high $E$.} For much smaller $E$ another dependence was detected
\begin{equation}\label{low_E}
Q\propto E^\alpha
\end{equation}
with $\alpha\approx 3$. Certainly, this dependence reflects the filling of traps with movable charges. Point of intersection of the two tangents to the linear regions of the $\log Q$ - $\log E$ plot may be used as an estimation for the total trapped charge $Q_t$ and, hence, to the total number of deep traps $M_0=Q_t/e$ or trap concentration $m_0=M_0/LS$ (here $L$ is the sample thickness and $S$ is the sample area). If this assumption is valid, then the intersection gives $m_0\approx 5\times 10^{13}$ cm$^{-3}$.

The major problem of this approach is a difficulty to motivate Eq. (\ref{low_E}) with $\alpha=3$ for the relation between $Q$ and $E$ in the low energy region. In the next section we consider a simple but realistic model of the trap-controlled carrier transport and suggest a more robust procedure for the determination of $m_0$. In the rest of the paper we mostly use the total number of the collected carriers $N=Q/e$ instead of the total extracted charge $Q$.

\section{MODEL OF TRAP-CONTROLLED CHARGE TRANSPORT}
\label{sec:model}

\subsection{No-recombination case}

Figure \ref{Q-E}b indicates that the total extracted charge is less than $CU$, hence, we can assume that the electric field in the sample is uniform in space and equal to the applied electric field $F$. This means that the carrier velocity $v$ is a constant in the bulk of the sample. As it was noted in Sec.~\ref{sec:exp}, we may also safely assume a spatially uniform initial distribution of the generated carriers.

Let us begin our consideration with the case when carrier
recombination is negligible. Then we can consider charges
of the opposite signs separately and limit our consideration to
the charges of one sign only, with concentration $n(x,t)$. Concentration of the empty traps is $m(x,t)$. Initial conditions are $n(x,0)=n_0$,
$m(x,0)=m_0$, so the total number of generated carriers is $N_0=n_0SL$. Dynamics of the model is governed by equations
\begin{eqnarray}
 \frac{\partial n}{\partial t}&=&
  -v\frac{\partial n}{\partial x}
  -knm, \label{no-recombination_eq1}
  \\
  \frac{\partial m}{\partial t}&=& -knm, \label{no-recombination_eq2}
\end{eqnarray}
which take into account the drift of charge carriers with average
velocity $v$ and a nonlinear term describing trapping of carriers
($k$ is the trapping rate constant). We do not take into account carrier diffusion, a brief motivation is provided in the Appendix.

If the initial distribution of carriers is uniform in
space, then drift does not produce a spatial variation of
 $n(x,t) $. The only effect of the drift is moving of the rear front of the carrier distribution with velocity $v$. Hence, for
$x > vt$ $n(x,t)=n(t)$ and we can solve Eqs. \eq{no-recombination_eq1} and \eq{no-recombination_eq2}
without the drift term, while for $x < vt$ $n(x,t)=0$ and
$m(x,t)=m(x,x/v)$.

\begin{figure}[bthp]
\begin{minipage}[c]{0.33\linewidth}
\begin{center}
\includegraphics[width=2in]{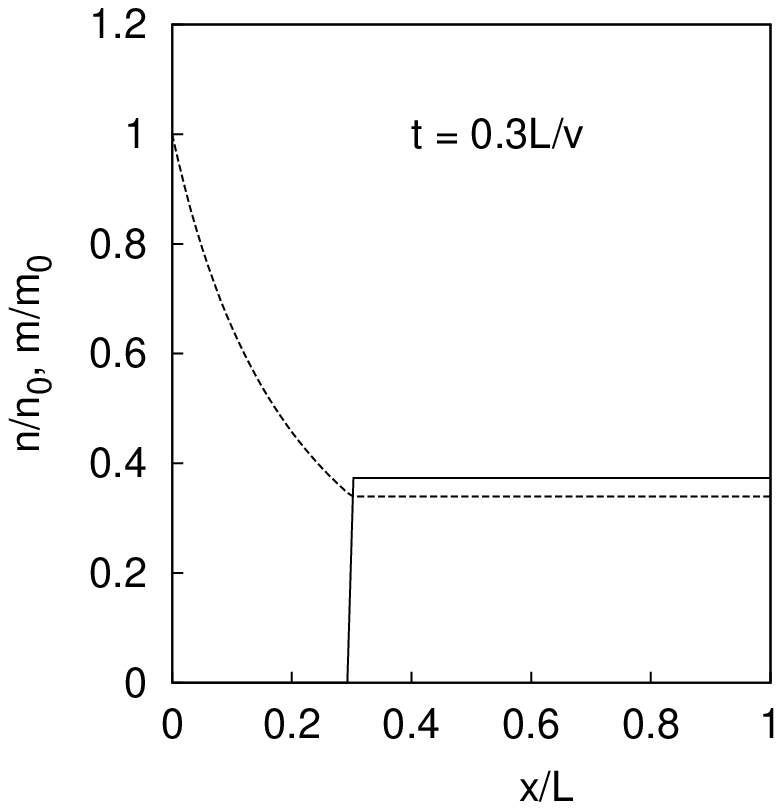}
\end{center}
\end{minipage}
\begin{minipage}[c]{0.33\linewidth}
\begin{center}
\includegraphics[width=2in]{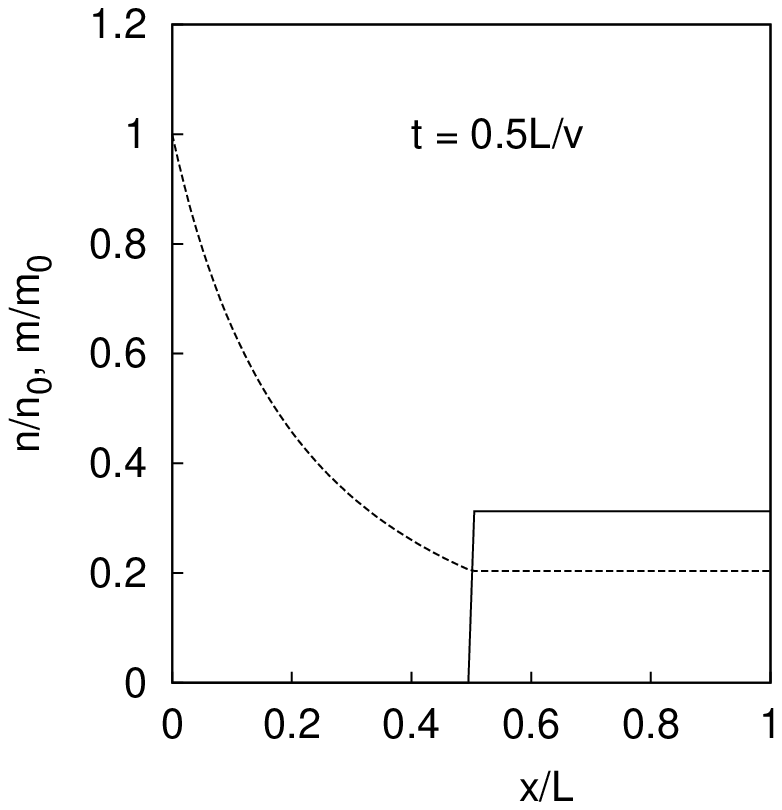}
\end{center}
\end{minipage}
\begin{minipage}[c]{0.33\linewidth}
\begin{center}
\includegraphics[width=2in]{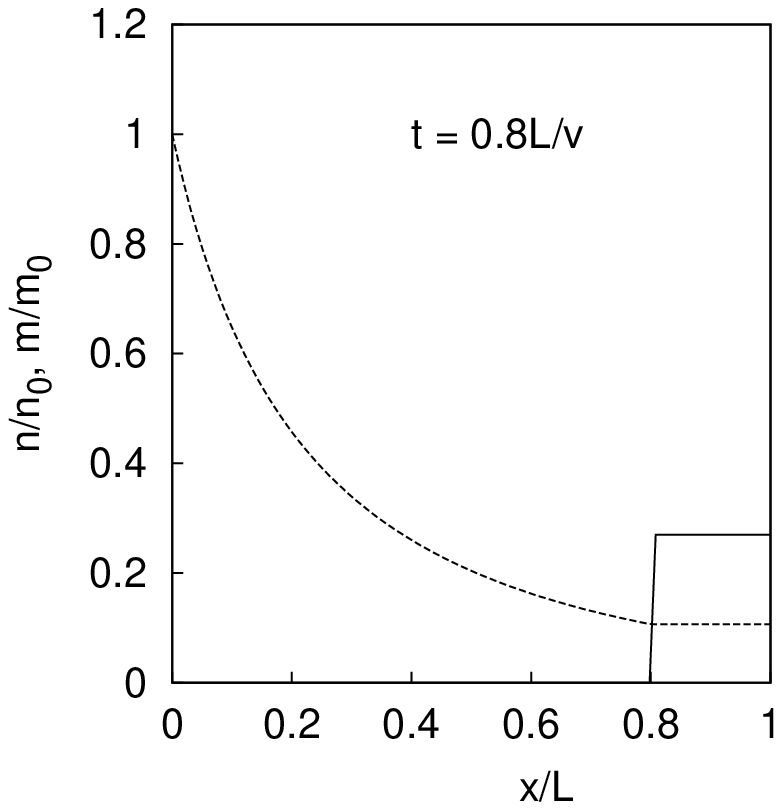}
\end{center}
\end{minipage}
\caption{Relative densities of free carriers $n(x,t)/n_0$ (solid lines) and empty traps $m(x,t)/m_0$ (broken lines) for three particular moments of time (shown in the plots); $q=1.5$, $\beta=0.3$.\label{distro}}
\end{figure}

Taking into account the conservation law
$n(x,t)-m(x,t)=n_0-m_0=\Delta$, we can write
\begin{equation}\label{no-recombination3}
  \frac{\partial n}{\partial t}=  -kn(n-\Delta), \hskip10pt
  \ln\frac{n_0(n-\Delta)}{n(n_0-\Delta)}=-\Delta k t, \hskip10pt
n(t)=\frac{n_0\Delta}{n_0-m_0\exp(-\Delta k t)}.
\end{equation}
Evolution of the carrier and empty trap distributions is shown in Fig. \ref{distro}.

\begin{figure}[thbp]
\begin{minipage}[c]{0.5\linewidth}
\begin{center}
\includegraphics[width=2.8in]{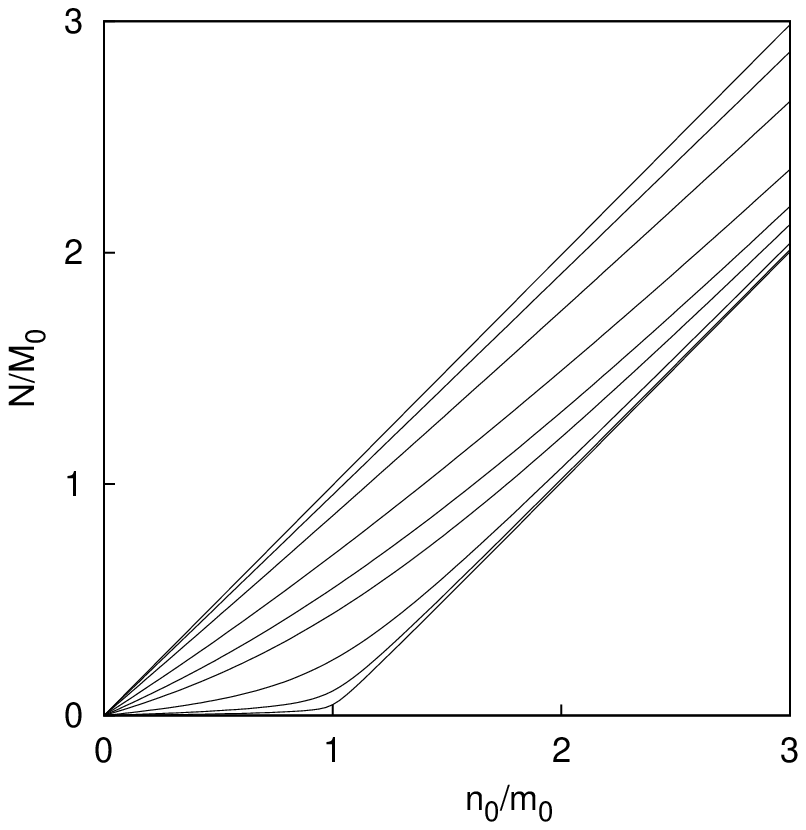}
\end{center}
\end{minipage}
\begin{minipage}[c]{0.5\linewidth}
\begin{center}
\includegraphics[width=2.8in]{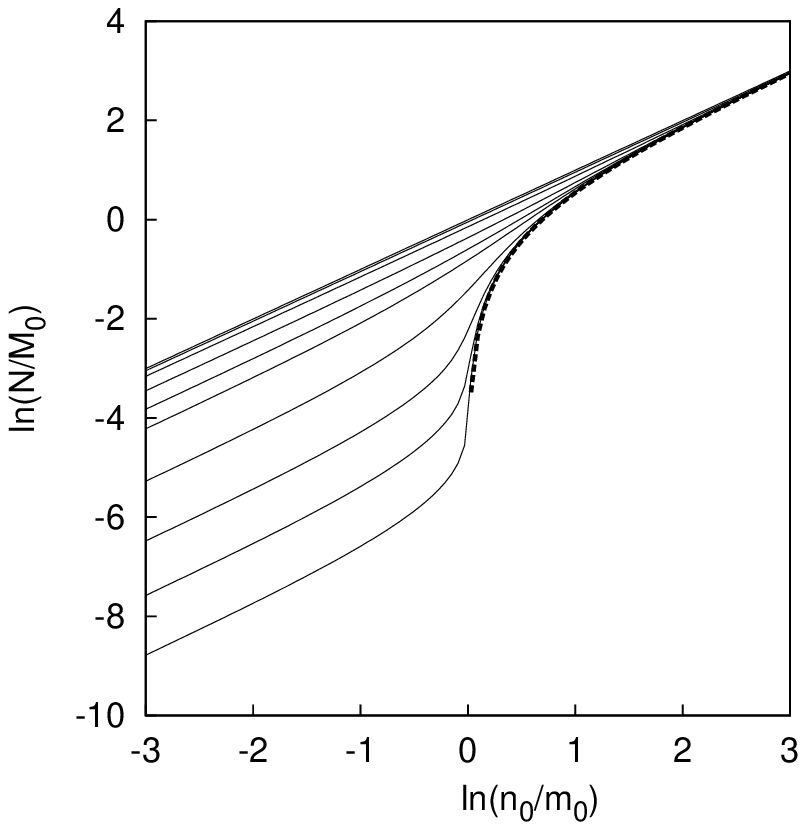}
\end{center}
\end{minipage}
\caption{General dependence of $N$ on $n_0$ in the double linear plot (left) and log-log plot (right) for different values
of $\beta$: 0.01, 0.03, 0.1, 0.3, 0.5, 1, 3, 10, and 100,
 from the bottom curve upward, correspondingly. Thick line is the limit
dependence $N/M_0=q-1$.\label{N(beta)}}
\end{figure}

The total number of extracted carriers is
\begin{equation}\label{total}
N=S\left[L\Delta+
  \int_0^L dx
  \left(n(x/v)-\Delta\right)\right]=S\left[L\Delta+\frac{v}{k}\ln\left(\frac{n_0-m_0
e^{-\Delta kL/v}}{\Delta} \right)\right].
\end{equation}
After suitable normalization
\begin{equation}\label{total_N}
  \frac{Q}{Q_t}=\frac{N}{M_0}=q-1 + \beta\ln\left(\frac{q-
e^{-(q-1)/\beta}}{q-1} \right), \hskip10pt
q=n_0/m_0=N_0/M_0, \hskip10pt \beta=\frac{v}{km_0L}.
\end{equation}
This formula is the main result of the paper. General behavior of
$N(n_0)$ is shown in Fig. \ref{N(beta)}. Taking into account that $N_0\propto E^2$, we see that Eq. \eq{total_N} gives the desired relation between $Q$ and $E$.\footnote{Strictly speaking, the relation $Q(E)$ should include contributions from the carriers of both signs, thus the more proper analogue of Eq. \eq{total_N} should have two separate terms with different parameters $q$ and $\beta$. In future we will not use this more strict relation because the limited accuracy of the experimental data gives no possibility for the reliable extraction of individual parameters $q$ and $\beta$ for positive and negative charges.}

Parameter $\beta=t_{\rm trap}/t_{\rm drift}$ is the ratio of
the characteristic trapping time $t_{\rm trap}=(km_0)^{-1}$ and
carrier drift time $t_{\rm drift}=L/v$. This helps to understand
most important features of Fig. \ref{N(beta)}. If $\beta \gg 1$,
then trapping is not effective and $N\approx N_0$. In the opposite
case of effective trapping $\beta \ll 1$
\begin{eqnarray}
 N&=&O(\beta), \hskip10pt n_0 < m_0, \label{eq1}
  \\
 N&=&N_0-M_0+o(1), \hskip10pt n_0 > m_0. \label{eq2}
\end{eqnarray}
In fact,  Eq. \eq{eq2} is a universal asymptotics of $N(N_0)$ for
$N_0\rightarrow \infty$ and any finite $\beta$, even for
$\beta\gg 1$, because for a very high concentration of carriers
all traps should be filled. Indeed, if $q \gg
\max(1,\beta)$, then
\begin{equation}\label{total_N_large_beta}
  \frac{N}{M_0}=q-1+\beta\left(\frac{1}{q}-\frac{1}{2q^2}\right)
  +O(\beta/q^3).
\end{equation}
For $\beta\gg 1$ this asymptotics develops only for a very high
concentration of carriers $q \gg \beta$ (for this reason it
is not seen in Fig. \ref{N(beta)} for $\beta \gg 1$).

The case of large $\beta$ (fast transport) is very unfavorable for the determination
of $m_0$. Thus, in such a case the experiment should be performed for
thick samples or a weak electric field (ensuring low $v$),
both conditions leading to the decrease of $\beta$.

For $q \ll 1$ the concentration of empty traps is approximately
constant, thus the only significant Eq. \eq{no-recombination_eq1} becomes linear. For this reason we must have a linear dependence $N=a(\beta)N_0$. This is indeed the
case
\begin{equation}\label{total_N_small_beta}
  \frac{N}{M_0}=\beta\left(1-e^{-1/\beta}\right)q +
  O(q^2).
\end{equation}

\subsection{Contribution from the carrier recombination}

Let us check, how sensitive are obtained results to the possible contribution from the
carrier recombination. In this case we have to consider two kinds of movable carriers
with concentrations $n(x,t)$ and $p(x,t)$. Dynamics of the model is described by equations
\begin{eqnarray}
 \frac{\partial n}{\partial t}&=&
  -v_n\frac{\partial n}{\partial x}
  -k_rnp, \label{recombination_eq1}
  \\
  \frac{\partial p}{\partial t}&=&v_p\frac{\partial p}{\partial x} -k_rnp, \label{recombination_eq2}
\end{eqnarray}
where $k_r$ is a recombination rate constant, and we omit trapping for the simplicity sake. Basic equations are
almost the same as in the previous case,\footnote{In fact, recombination kinetics may be considered as a special case of the trapping process for $q=1$.
For this case Eq. \eq{total_N} directly leads to Eq. \eq{recombination2b}
with $k_r=k$ and $v_p=0$.} but the reaction (recombination) zone is different because both species are movable. Again,
$n(x,t)=n(t)$ and $p(x,t)=p(t)$ for $v_nt < x < L-v_pt$.

Let us limit our consideration to the case $n_0=p_0$, so $n(t)=p(t)$ (if the
equality does not hold it implicitly means some sort of trapping
with trapped charges avoiding recombination). Temporal dependence
of $n(t)$ is
\begin{equation}\label{recombination2}
n(t)=\frac{n_0}{1+n_0k_rt}
\end{equation}
and the total number of carriers extracted at the collecting
electrode is
\begin{equation}\label{recombination2a}
\frac{N}{N_0}=1-\frac{1}{L}\int_0^{Lv_n/(v_n+v_p)} dx
\left(1-\frac{1}{1+n_0k_rx/v_n}\right)-\frac{1}{L}\int_0^{Lv_p/(v_n+v_p)} dx
\left(1-\frac{1}{1+n_0k_rx/v_p}\right)=
\end{equation}
\[
=\frac{v_n+v_p}{n_0k_rL}\ln\left(1+\frac{n_0k_rL}{v_n+v_p}\right).
\]
Finally,
\begin{equation}\label{recombination2b}
N=\eta\ln\left(1+\frac{N_0}{\eta}\right), \hskip10pt
\eta=\frac{(v_n+v_p)S}{k_r}.
\end{equation}

We conclude that the recombination is not important for $N_0 \ll \eta$, while
for $N_0\gg \eta$ it produces the dependence $N\propto \ln N_0$. Hence,
an experimental observation of the dependence $N\propto E^2\propto N_0$ for
high $E$ (and, hence, high initial number of carriers) is a clear indication that the recombination is negligible for the whole range of $E$.

\begin{figure}[htbp]
\begin{minipage}[c]{0.5\linewidth}
\begin{center}
\includegraphics[width=2.8in]{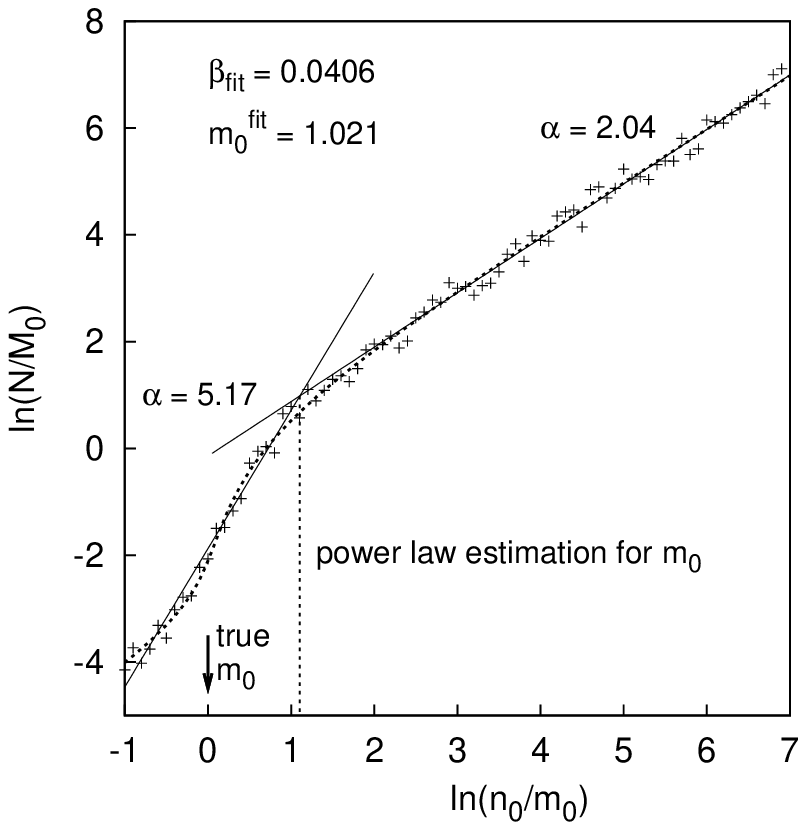}
\newline (a)
\medskip
\end{center}
\end{minipage}
\begin{minipage}[c]{0.5\linewidth}
\begin{center}
\includegraphics[width=2.8in]{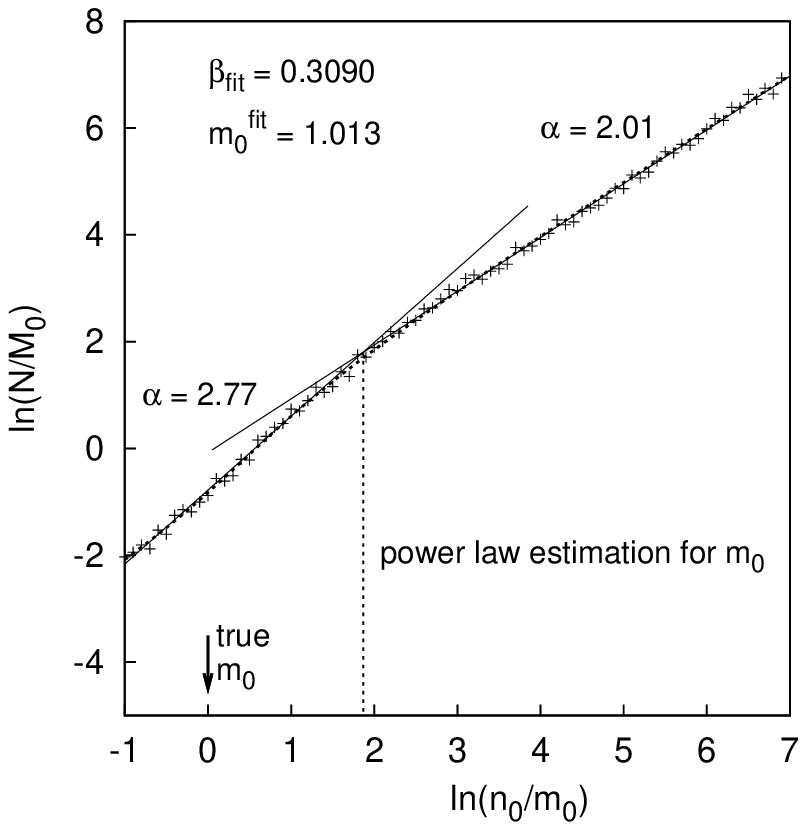}
\newline (b)
\medskip
\end{center}
\end{minipage}
\caption{Simulated data for $\beta=0.04$ (a) and $\beta=0.3$ (b), correspondingly, obtained by adding small random contributions to Eq. \eq{total_N}. We used $m_0=1$ in some arbitrary units in both plots. Plots show only a part of the full curve (for sufficiently high initial number of carriers), providing a more close resemblance with Fig. \ref{Q-E}b. Straight lines indicate the power law fit $N\propto n_0^{\alpha/2}\propto E^\alpha$, and broken lines show the result of the fit of the data to Eq. \eq{total_N}. Parameters $\beta_\textrm{fit}$ and $m_0^\textrm{fit}$, obtained by fitting, agree well with the true parameters.\label{simulate}}
\end{figure}

\subsection{Determination of a trap concentration from the experimental data}

Our consideration shows that for $N_0 \gg M_0$ there is a
natural relation
\begin{equation}\label{rel1}
N\approx N_0-M_0 \propto E^2-{\rm const},
\end{equation}
yet for
smaller $N_0$ there is a transition to another linear dependence
\begin{equation}\label{rel2}
N\approx a(\beta)N_0\propto E^2
\end{equation}
(see Fig. \ref{N(beta)}). Hence, in this model there is no dependence of the kind $N\propto E^\alpha$ with $\alpha > 2$. In fact, such dependence hardly exists in any model of the trap-controlled transport because the linear dependence of $N$ on $N_0$ for small concentration of carriers $n_0\ll m_0$ is a very general property of the charge transport, it just follows from the approximate constant concentration of empty traps in such a case. Nevertheless, the transition region from Eq. \eq{rel2} to Eq. \eq{rel1} can simulate this very kind of the dependence in a limit range of $E$ (see Fig. \ref{N(beta)}, the right plot).

If $\beta \ll 1$, then the concentration of traps could be estimated
by plotting of the experimental dependence $N$ vs $E^2$ and
drawing the linear asymptotics for the region of large $n_0$. If
$\beta \simeq 1$, then the most reliable way of the estimation
of $m_0$ is the direct fit of experimental data to Eq. \eq{total_N}.

Figure \ref{simulate} shows the result of fitting of a primitive simulation of the experimental data obtained as
\begin{equation}\label{simulation}
N^{\textrm{sim}}=N(n_0)(1+\delta),
\end{equation}
where $N(n_0)$ is calculated using Eq. \eq{total_N} and $\delta$ is a
random Gaussian number with zero mean and magnitude $\sigma=0.1$, which models a noise in the experimental data. One can see that the fit to the power law dependence \eq{low_E} produces strongly overestimated value of $m_0$, and the difference between $m_0^{\rm true}$ and $m_0^{\rm fit}$ progressively increases with the increase of $\beta$. Fit of the simulated data to Eq. \eq{total_N} produces a reasonable agreement.

\begin{figure}[bhtp]
\begin{minipage}[c]{0.5\linewidth}
\begin{center}
\includegraphics[width=2.8in]{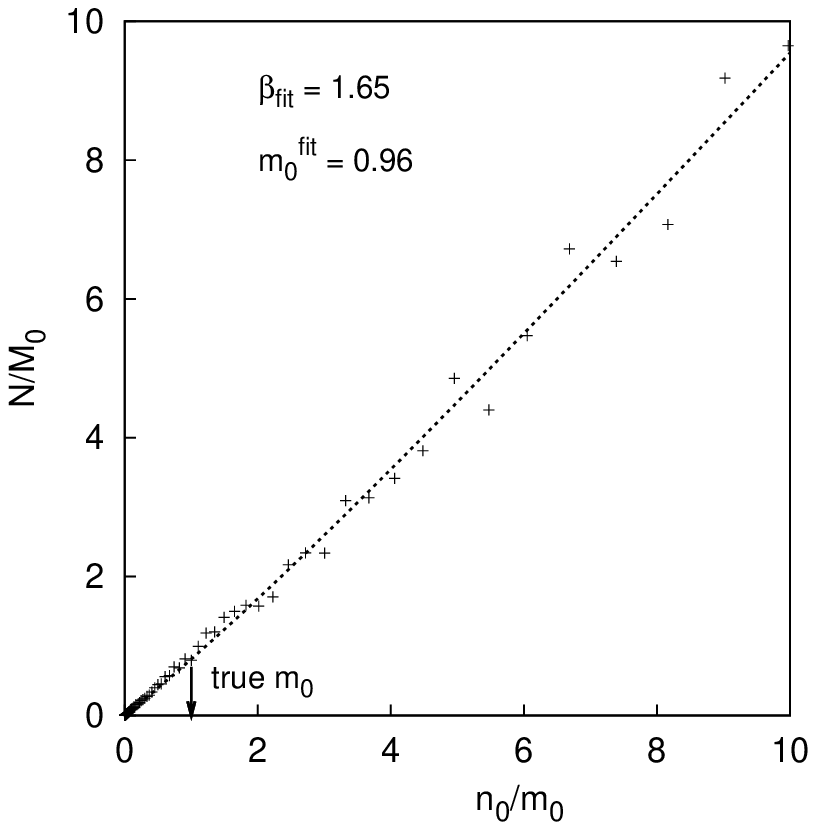}
\newline (a)
\medskip
\end{center}
\end{minipage}
\begin{minipage}[c]{0.5\linewidth}
\begin{center}
\includegraphics[width=2.8in]{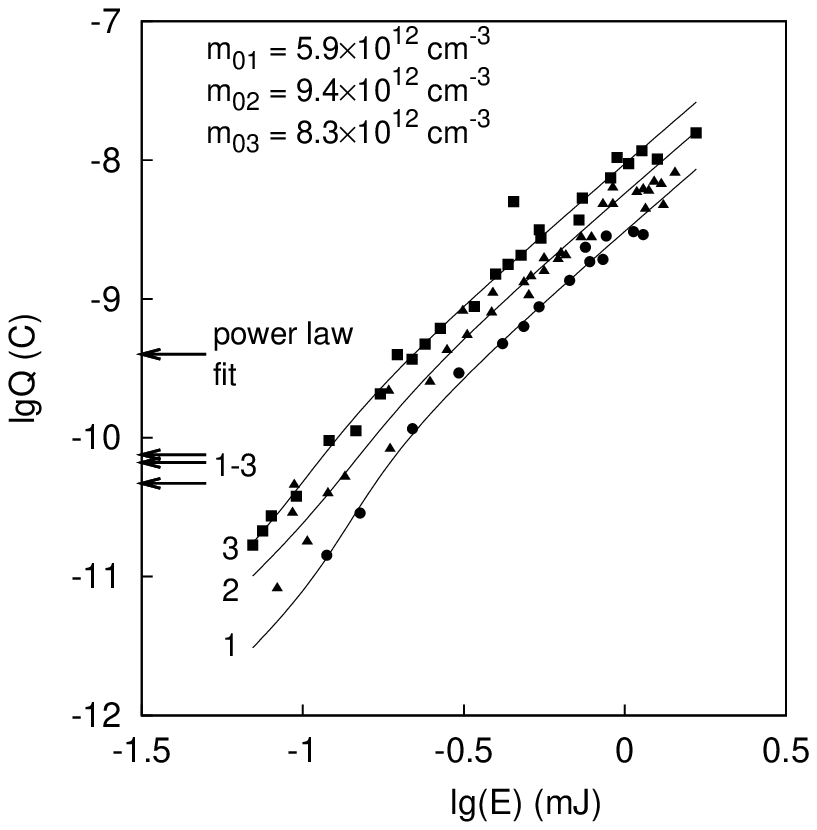}
\newline (b)
\medskip
\end{center}
\end{minipage}
\caption{(a) Simulated data for $\beta=3$. Broken line is the result
of the fit of the data to Eq. \eq{total_N}. (b) Fit of  experimental data for MEH-PPV, shown in Fig. \ref{Q-E}b, to Eq. \eq{total_N} (solid lines). Concentrations of deep traps, obtained by fitting, are shown in the plot, for different values of the applied voltage. These concentrations do not differ significantly, as it should be for the intrinsic parameter of the material. Some difference in values of $m_0$ for different values of $U$  is not surprising taking into account the poor accuracy of the experimental data.\label{simulate3}}
\end{figure}

Figure \ref{N(beta)} indicates that for $\beta \ge 1$ the reliable fit of
the experimental data to Eq. \eq{total_N} in the log-log plot is hardly possible. Reasonable question is: is it possible to estimate $m_0$ and $\beta$ in
such a case by fitting the data to Eq. \eq{total_N} in double linear coordinates? Figure \ref{simulate3}a shows that this is possible for $m_0$ if
experimental errors are not too large, but the estimation for
$\beta$ is still not reliable.

At last, we tried to fit the experimental data,
presented in Fig. \ref{Q-E}b, to Eq. \eq{total_N} (see Fig. \ref{simulate3}b). Again, true concentration of traps is much lower than the one, estimated
by the simple power law fit according to Eq. \eq{low_E}. It is worth to compare Figs. \ref{simulate}b and \ref{simulate3}b. In both cases for the low $E$ region $\alpha\approx 3$, and the corresponding ratios $m_0^{\rm power\hskip3pt law}/m_0^{\rm right\hskip3pt fit}$ are pretty close.


\section{discussion: when the method can be applied?}
\label{se:discussion}

Let us consider the applicability of the suggested method to real organic amorphous semiconductors. The necessary conditions for the direct applications of the method are: 1) spatially uniform generation of initial electron-hole pairs; 2) carrier recombination can be neglected; 3) the total trapped charge $Q_t=eM_0$ must be less that the extracted charge $Q=eN$ for the high end of the spanned range of $E$; 4) electric field is a constant in the bulk of the sample, so the total charge in the sample must be much less than the electrode charge $CU$, and 5) charge transport should not be very fast ($\beta \lesssim 1$).

First condition means in our case that the absorption of the light in the sample must be weak. The two-photon absorption as a way for the carrier generation is not important by itself for the application of the method. The only reason to use this particular method of the generation of initial carriers is a weak absorption of the 1064 nm light in MEH-PPV, thus ensuring the spatially uniform initial distribution of carriers. Other methods of carrier generation, capable to produce the uniform distribution, can be used as well.

In the second condition we mean a non-geminate recombination of charge carriers, because the geminate recombination could be phenomenologically taken into account simply by the redefinition of the proportionality coefficient $c$ in the relation between the light pulse energy and total number of initially generated free carriers $N_0=cE^2$. Absence of the non-geminate recombination can be reliably established by checking the linearity of the relation $Q\propto E^2$ for high $E$.

The third condition is a principal one, because if it is not valid, then we are dealing with the linear regime of low trap filling described by Eq. \eq{total_N_small_beta}, where the only relevant trap-related parameter is $km_0$ (see Eq. \eq{no-recombination_eq1} for the case $m(x,t)\approx m_0={\rm const}$). Naturally, the separate extraction of $k$ and $m_0$ is not possible in this regime. At the same time, this condition is not an obstacle \emph{per se} for the use of the method, it just dictates the proper range of $E$, most favorable for the determination of the trap density.

Last two conditions provide more serious restrictions because in some situations they are mutually contradictive. Indeed, the applied voltage $U$ should be high enough to provide the uniformity of the electric field in the sample, but at the same time the carrier velocity $v$  grows with the voltage, thus invariably shifting the sample to the unfavorable regime of high $\beta\propto v$. Yet the example of the MEH-PPV device indicates that for very typical organic materials the suggested method could be used successfully.

In the basic Eq. \eq{no-recombination_eq1} we omitted the diffusive term. Short analysis in the Appendix shows that the diffusive contribution to $N$ may be safely neglected in most cases.

\section{conclusion}
\label{se:conclusion}

In this paper we suggested the simple method for the estimation of the density of deep traps capable to keep charge carriers in amorphous semiconductors for a long time. One of the most important advantages of the method is a simplicity of the experimental set-up, which includes only standard widely used equipment. In the previous Section the major conditions, necessary for a successful application of the method, have been discussed. It turns out that these conditions are not very difficult to fulfil, thus the method may be considered as an almost universal one for the estimation of the density of deep traps in amorphous organic materials. Successful application of the method to the estimation of the concentration of deep traps in typical organic semiconductor MEH-PPV supports this conclusion.


\section*{APPENDIX. CONTRIBUTION FROM THE CARRIER DIFFUSION}
\label{app}

There is another limitation of the suggested method, directly related to the structure of our transport model: there is no diffusion term in Eq. \eq{no-recombination_eq1}, we set the diffusion coefficient $D=0$. Let us estimate when the diffusive contribution is negligible. Assuming $Lv/D \gg 1$ (weak diffusion), the diffusive contribution to $N$ could be estimated in the following way. At $t=0$ the distribution of movable carriers is uniform in space. It means that the diffusion provides the most important contribution at the rear front of $n(x,t)$, the only place where the gradient of $n$ is large. The width of the rear front increases with time as $\Delta_r(t)\simeq(2Dt)^{1/2}$, or, equivalently, with the traveled distance $x=vt$ as $\Delta_r(x)\simeq(2Dx/v)^{1/2}$. At the location of the rear front the concentrations of carriers and empty traps are, correspondingly
\begin{eqnarray}
 n_r(x)&=&\frac{n_0(n_0-m_0)}{n_0-m_0\exp\left[-kx(n_0-m_0)/v\right]}, \label{nr}
  \\
  m_r(x)&=&n_r(x)-n_0+m_0. \label{mr}
\end{eqnarray}
Diffusive addition $\delta N$ to the total $N$ can be estimated as a number of carriers, trapped in the diffusive zone with width $\Delta_r$, moving with the rear front of the carrier distribution
\begin{equation}\label{addN}
\delta N\simeq\frac{kS}{v}\int_0^L dx \Delta_r(x)n_r(x)m_r(x).
\end{equation}
We neglect here a variation of $n(x,t)$ in the vicinity of the rear front, this changes the estimation \eq{addN} by the factor $\simeq O(1)$. Parameter $\Delta_r$ varies slowly with $x$, it can be estimated as $\Delta_r(L)$ and taken out of the integral. Next, we consider only two limiting cases $n_0 \gg m_0$ and $n_0 \ll m_0$. If $n_0 \gg m_0$, then
\begin{eqnarray}
 n_r(x)&\simeq&n_0, \nonumber
  \\
  m_r(x)&\simeq&m_0\exp\left(-kxn_0/v\right), \nonumber
  \\
  \frac{\delta N}{N}&\simeq&\frac{m_0}{n_0}\left(\frac{D}{Lv}\right)^{1/2}\left[1-\exp\left(-kLn_0/v\right)\right]. \label{dn2}
\end{eqnarray}
Estimation \eq{dn2} is indeed small. For the opposite case $m_0 \gg n_0$
\begin{eqnarray}
 n_r(x)&\simeq&n_0\exp\left(-kxm_0/v\right), \nonumber
  \\
  m_r(x)&\simeq&m_0, \nonumber
  \\
  \frac{\delta N}{N}&\simeq&\frac{1}{\beta}\left(\frac{D}{Lv}\right)^{1/2}. \label{dn3}
\end{eqnarray}
To obtain Eq. \eq{dn3} we used Eq. \eq{total_N_small_beta}, which relates $N$ with $N_0$ for $m_0\gg n_0$. For $\beta\simeq 1$ this is a small contribution, but it can be large if $\beta \ll 1$. Hence, in some cases of a slow charge transport the ratio $\delta N/N$ may be not negligible, but this is most probable for a very far left end of the spanned range of $E$, because the factor $\left(D/Lv\right)^{1/2}$ is typically small in amorphous organic materials at the room temperature.

This factor may be estimated using the Einstein relation $\mu=eD/kT$. This relation is not strictly valid in organic amorphous materials,\cite{hirao,parris,novikov} but could be used as a very crude estimation of $D$. According to the Einstein relation $D/Lv\simeq kT/eU$. This factor is indeed very small for the typical applied voltage $U\simeq 10\div 100$ V and the room temperature with $kT=0.026$ eV. We conclude that the typical diffusive contribution to Eq. \eq{total_N} is not very important.

\acknowledgments     

This work was partly supported by the Russian Foundation for Basic Research (grants  08-03-00125 and 10-03-92005) and the International Science and Technology Center (grant  3718).


\end{document}